\documentclass[twocolumn]{revtex4}
\usepackage{graphicx}
\usepackage{scrtime}
\usepackage{latexsym}
\usepackage{epsfig}
\usepackage{fancyhdr}
\usepackage[hang]{subfigure}
\usepackage{supertabular}
\usepackage{wrapfig}
\usepackage{amssymb}

\begin{document}


\title{Modulating the growth conditions:  Si as an acceptor in (110) GaAs for high mobility p-type heterostructures}

\author{F. Fischer}
\author{M. Grayson}
\author{D. Schuh}
\author{M. Bichler}
\author{G. Abstreiter}
\affiliation{Walter Schottky Institut, Technische Universit\"at M\"unchen, 85748, Garching, Germany}
\author{K. Neumaier}
\affiliation{Walther Meissner Institut, Bayerische Akademie der Wissenschaften, 85748 Garching, Germany}

 
\begin{abstract}

We implement metallic layers of Si-doped (110) GaAs as modulation doping in high mobility p-type heterostructures, changing to p-growth conditions for the doping layer alone.  
The strongly auto-compensated doping is first characterized in bulk samples, identifying the metal-insulator transition density and confirming classic hopping conduction in the insulating regime.  
To overcome the poor morphology inherent to Si p-type (110) growth, heterostructures are fabricated with only the modulation doping layer grown under p-type conditions.  
Such heterostructures show a hole mobility of $\mu = 1.75\times10^5~ \textnormal{cm}^2/\textnormal{V}\textnormal{s}$ at density $p=2.4\times10^{11}~ \textnormal{cm}^{-2}$.  
We identify the zero field spin-splitting characteristic of p-type heterostructures, but observe a remarkably isotropic mobility and a persistent photoconductivity unusual for p-heterojunctions grown using other doping techniques.  
This new modulated growth technique is particularly relevant for p-type cleaved-edge overgrowth and for III-V growth chambers where Si is the only dopant.

\end{abstract}


\maketitle

\begin{figure}
\center
\includegraphics[width=7cm,keepaspectratio]{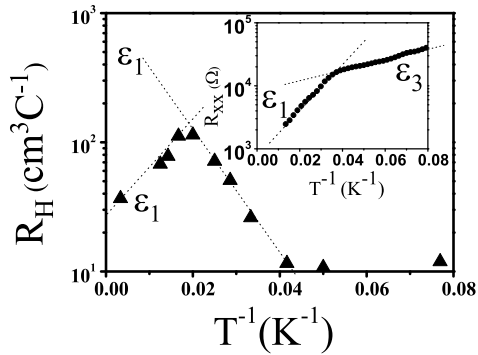}
\caption{Arrhenius plot of Hall coefficient showing a non-monotonic temperature dependence. 
The dotted lines indicate the associated activation energies. Inset: longitundinal resistance exhibits activated behavior}
\label{hall_coefficient}
\end{figure}

\begin{figure}
\center
\includegraphics[width=9cm,keepaspectratio]{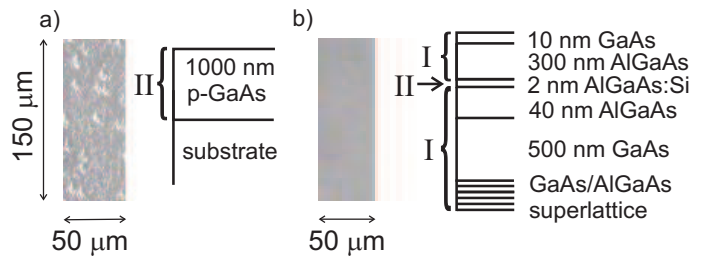}
\caption{Microscopic picture of a surface and the corresponding growth protocol of a) a $1~\mu m$ thick p-type layer (left) and b) a 2DHG sample of equivalent thickness using modulated growth conditions(right)}
\label{microscope_surface}
\end{figure}

\begin{figure}
\center
\includegraphics[width=8cm,keepaspectratio]{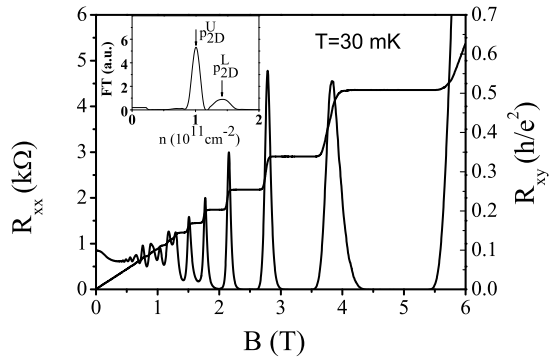}
\caption{Longitudinal and quantum Hall resistance data of 2D holes in (110) GaAs. The inset shows the shows the a Fourier transformation of the longitudinal resistance in the low B-field region.}
\label{sample_sdh_data}
\end{figure}

\indent High mobility two-dimensional (2D) holes in GaAs are of research interest both for their strong spin-orbit coupling and for their heavy mass.  
Studies of spin effects may have implications for the field of spintronics, since a large Rashba term splits the spin subbands even at zero external magnetic field, $B$ \cite{ winkler:1}.  
The large mass has lead to studies of a new kind of metal insulator transition and evidence of a Wigner crystal \cite{ papadakis:5592}.\\
\indent Previously 2D hole gas (2DHG) systems 
have been grown on different facets of GaAs substrates using Be \cite{stormer:685} or C \cite{wieck:51} as an acceptor. Si was used as an acceptor either
on $\left(311\right)\textnormal{A}$ substrate where the highest 2DHG hole mobilities have been achieved. 
The disadvantage of Be is that 
its high vapor pressure and diffusivity causes it to contaminate the growth chamber, and 
introducing it into a high mobility III-V growth chamber leads to a general degradation of sample quality.\\
\indent In this Letter, we present a new p-type modulation doping method for (110) GaAs whereby the growth conditions for the Si modulation doping alone are adjusted to induce incorporation as an acceptor.
We first confirm the bulk metal-insulator transition (MIT) under high autocompensation, complemented with a study of the activated hopping in the insulating state, and then we use the metallic layers as modulation doping for a high mobility p-type heterojunction. \\
\indent We begin with 
studies of the MIT 
in bulk p-type doped samples. 
To calibrate the Si doping density and efficiency for p-type growth on $\left(110\right)$ GaAs, we grew 
several bulk doped $1~\mu\textnormal{m}$ thick, non-degenerately and degenerately p-doped samples with different doping densities. 
Si is amphoteric in GaAs 
and can act as both donor and acceptor during growth depending on its incorporation on the Ga or As lattice site.
Especially at morphology optimized growth conditions on the $\left(110\right)$ GaAs surface, 
labeled Type I 
(Table I), Si acts 
predominantly 
as donor.
\begin{table}
\center
\begin{tabular}{|c|c|c|c|c|c|}
\hline
  & & & & $\textnormal{GaAs}$  & \\
  & & \raisebox{1.5ex}[-1.5ex]{$ \textnormal{T}$}  & \raisebox{1.5ex}[-1.5ex]{$\textnormal{P}^{\textnormal{BEP}}_{\textnormal{As}_4}$} & $\textnormal{growth rate}$ & \raisebox{1.5ex}[-1.5ex]{$\textnormal{morphology}$}\\
\hline
I & n-type & $490^\circ\textnormal{C}$           & $6\times10^{-5}~\textnormal{mbar}$          & $0.177~\textnormal{nm}/\textnormal{s}$ & $\textnormal{excellent}$ \\
\hline
II & p-type & $640^\circ\textnormal{C}$           & $1\times10^{-5}~\textnormal{mbar}$          & $0.177~\textnormal{nm}/\textnormal{s}$ & $\textnormal{poor}$ \\
\hline
\end{tabular}
\label{growth_conditions}
\caption{Growth conditions for n-type (type 1) and p-type (type II) bulk growth}
\end{table}
However, when the substrate temperature is increased and the As pressure lowered,
labelled Type II 
(Table I), the number of Si-acceptors outweighs the Si-donors leading to a net p-doping.
Growing a thick p-doped layer under such conditions, resulted in a surface covered with large, triangular-shaped defects (Fig. \ref{microscope_surface}a) as observed 
previously
\cite{tok:4160}.\\
\indent We define the acceptor efficiency $\eta$ as the ratio of Si-acceptors $N_A$ to the total number of incorporated silicon atoms $N_{Si}$ and the compensation ratio $\kappa$ as the ratio of Si-acceptors $N_A$ to Si donors $N_D$. We calibrate assuming a
Si-acceptor 
doping efficiency of $\eta = N_A/N_{Si}=0$ 
$\left( \kappa = N_A/N_D=0 \right)$ for n-type GaAs on a $\left( 001\right)$ oriented substrate. 
%
Samples with an acceptor density increased to $N_A=7.1\times10^{18}~\textnormal{cm}^{-3}$ exhibit a metallic behavior down to 4.2 K.
The hole density is $p=2.1\times10^{18}\textnormal{cm}^{-3}$ with $\eta_{4.2K} \approx 0.59$ $\left( \kappa_{4.2K} \approx 1.45 \right)$ for $N_{Si} = 1.2\times10^{19}\textnormal{cm}^{-3}$.
The low density sample has $p=1.7\times10^{17}\textnormal{cm}^{-3}$ at $300~\textnormal{K}$. 
With $p=N_A-N_D$ at room temperature one deduces $\eta_{300K} \approx 0.55$ 
$\left( \kappa_{300K} \approx 1.22 \right)$ 
at a Si density of $N_{Si} = 1.7\times10^{18}\textnormal{cm}^{-3}$ at $300~\textnormal{K}$.\\
\indent Transport in the heavily auto-compensated insulating sample is well-described by hopping conduction in a parallel impurity band. 
The temperature dependence of the Hall coefficient of this sample (Fig. \ref{hall_coefficient}) 
is expected to show the following functional form (following the notation of \cite{ efros:1}): 

\begin{displaymath}
R_H(T) \sim \left\{
\begin{array}{ll}
e^{\epsilon_1/kT} & T>50~\textnormal{K} \\
e^{-\epsilon_1/kT} & 25~\textnormal{K}<T< 50~\textnormal{K}  \\
\textnormal{const} &   T< 25~\textnormal{K}
\end{array}
\right.
\end{displaymath}
with the longitudinal resistance (inset Fig. \ref{hall_coefficient})
 $R_{xx}\sim e^{\epsilon_3/kT}$ for $T<28~\textnormal{K}$ and $R_{xx}\sim e^{\epsilon_1/kT}$ for $T>28~\textnormal{K}$.
This is typical for lightly bulk-doped semiconductors with conduction in the valence band and in an impurity band, with the activated behavior $\epsilon_3= 2.1~\textnormal{meV}$ characterizing the nearest neighbor hopping gap \cite{efros:1}.
The deduced acceptor activation energy $\epsilon_1 = 7.8~\textnormal{meV}$
is about a factor of $\times5$ smaller than the activation energy $\Delta E = 35~\textnormal{meV}$ of isolated Si-acceptors in GaAs obtained by Hall-effect measurements \cite{sze:1}. Taking into account that the activation energy in doped semiconductors decreases according to $\epsilon_1 = \Delta E \left[ 1 - (N_A/N_{crit})^{1/3}\right]\approx 14~\textnormal{meV}$ \cite{schubert:1}, the measured $\epsilon_3$ is lower than the expected one 
and may be 
due to an impurity band strongly broadened towards the valence band.
From the value of $R_H$ in the saturation regime for low temperatures where the valance band holes are frozen out, one can deduce the number of hopping conductors $p_{N_A}=N_A-N_D=5.2\times10^{17}~\textnormal{cm}^{-3}$ involved in transport,
leading to $\eta_{<25K} \approx 0.65$ 
$\left( \kappa_{<25K} \approx 1.86 \right)$ 
  A discrepancy between $\eta_{<25K}$ and $\eta_{300K}$ has been also observed 
previously \cite{arushanov:2653} 
with the former being relevant for low-temperature applications. \\ 
\indent Now we can compare the experimental results with the theoretical expectations for the MIT.  
The metal-insulator transition is defined by the Mott criterion $a^*_B N_{crit}^{1/3} \approx 0.25$, with $a_B^*\approx15~\textnormal{\AA}$ the effective Bohr radius and $N_{crit}$ the critical transition density \cite{schubert:1} which can be estimated to $N_{crit} 
= 
5\times10^{18}~\textnormal{cm}^{-3}$ for acceptors in p-type GaAs. The deduced acceptor concentration $N_A=1.1\times10^{18}~\textnormal{cm}^{-3} < N_{crit}$ 
showing insulating behavior is well below the critical density of the MIT. 
In 
agreement 
with the theory the density $N_A=7.1\times10^{18}~\textnormal{cm}^{-3} > N_{crit}$ 
showing metallic behavior is above the critical density of the MIT.\\
\indent The rest of the article discusses the two-dimensional hole gas, which was fabricated by using a thin layer of the metallic p-type material as a modulation doping.
To overcome the problem of the poor surface morphology and the consequent bad mobility, we grew the modulation doped heterostructure by using two different growth conditions.
We used type I growth conditions 
up to the doping layer. Then we changed to type II for the doping and switched back to type I conditions (Fig. \ref{microscope_surface}b).
This resulted in a very good surface morphology for the 2DHG growth (Fig. \ref{microscope_surface}b), comparable to 
2DEG heterostructures grown on $\left(110\right)$ GaAs \cite{fischer:108}.\\
\indent 
We studied the sample quality with measurements of the longitudinal and 
transverse 
resistance with a special focus on the low B-field Shubnikov-de Haas oscillations (SdH).
The highest observed hole mobility is $ \mu = 175~000~\textnormal{cm}^2/\textnormal{Vs}$ at a density of $p_{2D} = 2.4 \times 10^{11} \textnormal{cm}^{-2}$ indicating the excellent quality of the samples despite the doping layer with the poor morphology. An ungated sample shows very well developed quantum Hall plateaus and minima of the longitudinal resistance at $30~\textnormal{mK}$ (Fig. \ref{sample_sdh_data}). In the low B-Field regime one can observe a beating of the SdH (Fig. \ref{sample_sdh_data}) which is known from literature for 2DHGs \cite{stormer:126} 
to arise 
from zero-field spin splitting because of inversion asymmetry \cite{winkler:1}. By plotting $R_{xx}$ against $1/B$ and Fourier transforming it \cite{habib:113311} in the low B-Field region (inset Fig. \ref{sample_sdh_data}) one can directly deduce the density of $p^L_{2D} = 1.4 \times 10^{11} \textnormal{cm}^{-2}$ of the lower spin subband and the density  $p_{2D}^U = 1.0 \times 10^{11} \textnormal{cm}^{-2}$ of the upper subband which add up to the total density evident at high B-fields $p_{2D} = 2.4 \times 10^{11} \textnormal{cm}^{-2}$.

\begin{figure}
\center
\includegraphics[width=8cm,keepaspectratio]{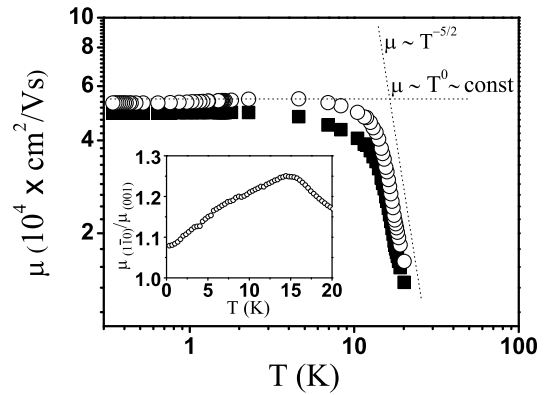}
\caption{Log-log plot of temperature dependent in-plane mobility anisotropy for the $\left[1\overline{1}0\right]$ ($\circ$) and $\left[001\right]$ ($\blacksquare$) direction. 
The two principal power-laws are indicated by the dotted lines. Inset: temperature dependent ratio of the 
two mobilities.}
\label{temp_dep_mob_aniso}
\end{figure}

We studied persistent photoconductivity (PPC) which is a common technique to increase the carrier density in two-dimensional electron systems (2DES) 
though 
it is uncommon in 2DHGs \cite{stoermer:139}. 
We find a pronounced PPC effect in our samples.
The unilluminated, dark density of the 2DHG at $4.2~\textnormal{K}$ is $p_{2D} = 1.17 \times 10^{11} \textnormal{cm}^{-2}$ with a mobility of 
$ \mu = 28~000~\textnormal{cm}^2/\textnormal{Vs}$. By illuminating a Hall-bar sample with a LED we were able to tune the density up to the maximum density of $p_{2D} = 2.3 \times 10^{11} /\textnormal{cm}^{-2}$ with a mobility 
$\mu = 140~000~\textnormal{cm}^2/\textnormal{Vs}$ at $4.2~\textnormal{K}$.\\
\indent By using an L-shaped Hall-bar we investigate the mobility anisotropy of the two principal in-plane directions
$\left[1\overline{1}0\right]$ and $\left[001\right]$.
A considerable mobility anisotropy has been reported in literature for
2DHGs in GaAs \cite{henini:451}. Up to a factor of 4 anisotropy has been observed in Si-doped $\left(311\right)\textnormal{A}$ systems and different Be-doped facets and is partially due to anisotropic interface roughness scattering \cite{heremans:1980} and the anisotropic effective mass. 
Henini, et al. \cite{henini:446} found that in Be-doped $\left(110\right)$ 2DHGs the mobility along $\left[1\overline{1}0\right]$ is about a factor of 3 larger than the mobility along $\left[001\right]$ at a density of $p_{2D} = 1.79 \times 10^{11} \textnormal{cm}^{-2}$. In contrast, in our samples we observed only a very weak anisotropy 
mostly pronounced at low temperatures where it is within 
a factor of 
$1.1$ to $1.3$. 
Fig. \ref{temp_dep_mob_aniso} shows that the mobility is nearly constant for $T<10~\textnormal{K}$ in both directions and that for $T>10{\textnormal{K}}$, $\mu \sim T^{-5/2}$. 
From anisotropies in the bulk mass alone, one would expect a mobility anisotropy of a factor of 0.5, leaving open the possibility that anisotropic roughness may be counteracting the mass anisotropy.\\
\indent In summary we 
characterized heavily autocompensated Si-doped bulk (110) GaAs, and 
observed an insulating impurity band for a non-degenerate 
bulk p-type sample below the MIT, and metallic conduction for a degeneratly doped sample. We were able to grow the first high mobility 2DHG on $\left(110\right)$ oriented GaAs using Si as a dopant. We observed an unusual PPC and a weak anisotropy. \\  
\indent High mobility MBE systems, equipped with Si as the {\it only} dopant source can take advantage of this new technique to grow 2DHGs on $\left(110\right)$ oriented GaAs. Since the $\left[110\right]$ orientation is the natural cleavage plane of GaAs, it should also be possible to grow 2DHGs by cleaved edge overgrowth (CEO) \cite{pfeiffer:1697} without introducing additional dopant sources. This enables new devices like atomically precise hole wires and orthogonal 2D-2D tunnel-junctions, for probing the quantum Hall edges of a 2DHG analogous to previous studies in electron systems \cite{huber:164}. It also permits bipolar n-p heterostructures to be prepared in standard III-V growth chambers.\\
\begin{acknowledgements}
This work was supported financially by Deutsche Forschungsgemeinschaft via Schwerpunktprogramm Quantum-Hall-Systeme
and in the framework of the COLLECT EC-Research Training Network HPRN-CT-2002-00291. 
\end{acknowledgements}

\end{document}